\documentstyle[sprocl]{article}
\def\Journal#1#2#3#4#5#6{{#1} {\bf #2}, #3 (#5)}

\def\CQG{\em Class. Quantum Grav.}

\def\JMP{\em J. Math. Phys.}

\def\AIHP{\em Ann. Inst. H. Poincar\'e}

\def\CR{\em C.R. Acad. Sci. (Paris)}

\newcommand{\bm}[1]{\mbox{\boldmath $#1$}}
\def\d{{\rm d}}
\def\g{{\rm g}}
\def\lie{{\cal L}}
\def\espaitemps{({\cal V},\g)}
\def\varietat{{\cal V}}

\def\CT#1#2{{\cal J}^{}_{#2}\left(\vec #1\right)}
\def\CB#1#2#3#4#5{j^{#5}_{#4}\left( R_{[2],[2]};\vec{#1},\vec{#2},\vec{#3}\right)}

\def\be{\begin{equation}}
\def\ee{\end{equation}}
\def\bea{\begin{eqnarray}}
\def\eea{\end{eqnarray}}
\def\bean{\begin{eqnarray*}}
\def\eean{\end{eqnarray*}}

\newcounter{eqlletra}

\newtheorem{theorem}{Theorem}[section]

\newtheorem{corollary}{Corollary}[theorem]

\begin{document}

\title{Properties of Bel currents}
\author{R. Lazkoz}
\address{Facultad de Ingenier\'{\i}a (ESIDE), Universidad de Deusto,
Apdo. de Correos, 1, 48080 Bilbao, Spain}
\author{J.M.M. Senovilla}
\address{Fisika Teorikoaren Saila, Euskal Herriko Unibertsitatea,
644 P.K., 48080 Bilbao, Spain}
\author{R. Vera}
\address{School of Mathematical Sciences, Queen Mary University of London,
Mile End Road, London E1 4NS, England, UK}

\maketitle\abstracts{The Bel tensor is divergence-free in some
important cases leading to the existence of conserved currents associated
to Killing vectors analogously to those of the energy-momentum tensor.
When the divergence of the Bel tensor does not vanish one can study the
interchange of some quantities between the gravitational and other
fields obtaining mixed total conserved currents. Nevertheless, the
Bel currents are shown to be conserved (independently of the matter
content) if the Killing vectors satisfy some very general
conditions. These properties are similar to some very well known
statements for the energy-momentum tensor.}

\section{Introduction}
A consequence of the Principle of Equivalence is that
there is no possible proper definition of a {\em local energy-momentum tensor}
for the gravitational field, where by ``proper'' it is meant a
tensor constructed from the metric and its first derivatives.
Nevertheless, there exist {\em local} tensors describing the {\em strengh}
of the gravitational field. The outstanding example is the
so called Bel--Robinson (B-R) tensor \cite{B4,Beltesis}, a four-index tensor
constructed for vacuum spacetimes whose properties are similar to the
traditional energy-momentum (e-m) tensors:
it has analogous positivity properties, it vanishes
iff the curvature does, it is divergence free, and others.
But it is not an energy-momentum tensor: it has four indices, and
its physical dimensions are not those of an energy density (it was
Bel himself who refer to that dimensions as `super-energy', whose
best interpretation so far is that of energy per surface unit).
Therefore, instead of trying to define a sort of `energy' for the
gravitational field, the idea behind the works on super-energy
(s-e) tensors is to define analogous objects for the rest of the physical
fields which ought to be related somehow at this s-e level.
A purely algebraic construction of s-e tensors for arbitrary fields
was presented in \cite{jose.s-e}, and includes the usual Bel tensor,
which generalizes the B-R tensor for non-vacuum spacetimes.
We refer to \cite{contri} for a discussion on this issue.
Contracting the Bel tensor with Killing vectors one constructs
some `Bel currents' 
which are not divergence-free in
general (the matter acts as source), so that they may
lead to the interchange of ``s-e'' quantities between the
gravitational and other physical
fields. This possibility is analyzed in \cite{jose.s-e}
for the cases of the minimally coupling with a scalar field,
electromagnetic and Proca fields,
using as inspiration the \underline{mixed} divergence-free currents
(which are not conserved separately) traditionally found by adding
the e-m tensors describing fields in interaction.

This s-e interchange has been found in spacetimes
admitting a non orthogonally transitive $G_2$ group of isometries,
in the Wils' family of stiff fluid solutions \cite{wils1}, see \cite{rjr}.
In simpler cases the Bel currents are shown \cite{rjr} to be
conserved automatically depending on some geometrical properties of the
Killing vectors used in their construction, independently of the matter
content \cite{rjr}.
These properties are the analogous to some very well known statements
concerning the e-m tensors. 

The covariant derivative associated to $\g$ in our spacetime
$\espaitemps$ will be denoted both by $\nabla$ and $;$.
Round and square brakets embracing any number of indices will denote the
usual symmetrization and antisymmetrization, respectively.
Greek indices run from $0$ to $3$.

\section{The Bel tensor and its currents}
The Bel tensor \cite{B4,Beltesis},
which can be thought as being
the basic super-energy tensor ($T$) for the gravitational
field constructed with the Riemann tensor $R_{\alpha\beta\lambda\mu}$
(double symmetric (2,2)-form, i.e. $R_{[2],[2]}$)
in four dimensions \cite{jose.s-e}, reads as follows
\begin{eqnarray}
B_{\alpha\beta\lambda\mu}\equiv 
T_{\alpha\beta\lambda\mu}\left\{R_{[2],[2]}\right\}=
R_{\alpha\rho\lambda\sigma}
R_{\beta}{}^{\rho}{}_{\mu}{}^{\sigma}
+R_{\alpha\rho\mu\sigma}
R_{\beta}{}^{\rho}{}_{\lambda}{}^{\sigma}-\nonumber\\
-\frac{1}{2}g_{\alpha\beta}
R_{\rho\tau\lambda\sigma}R^{\rho\tau}{}_{\mu}{}^{\sigma}
-\frac{1}{2}g_{\lambda\mu}
R_{\alpha\rho,\sigma\tau}R_{\beta}{}^{\rho\sigma\tau}+
\frac{1}{8}g_{\alpha\beta}g_{\lambda\mu}
R_{\rho\tau\sigma\nu}
R^{\rho\tau\sigma\nu},\nonumber
\end{eqnarray}
from where the following symmetry properties for $B$ explicitly arise
\begin{equation}
B_{\alpha\beta\lambda\mu}=B_{(\alpha\beta)(\lambda\mu)}=
B_{\lambda\mu\alpha\beta}.
\label{belsym}
\end{equation}

Using the second Bianchi identity $\nabla_{[\nu}R_{\alpha\beta]\lambda\mu}=0$
one obtains the following expression for the divergence of the Bel tensor
\begin{equation}
\nabla_{\alpha}B^{\alpha\beta\lambda\mu}=
R^{\beta\hspace{1mm}\lambda}_{\hspace{1mm}\rho\hspace{2mm}\sigma}
J^{\mu\sigma\rho}+R^{\beta\hspace{1mm}\mu}_{\hspace{1mm}\rho\hspace{2mm}\sigma}
J^{\lambda\sigma\rho}-\frac{1}{2}g^{\lambda\mu}
R^{\beta}_{\hspace{1mm}\rho\sigma\gamma}J^{\sigma\gamma\rho},
\label{divbel}
\end{equation}
where $J_{\lambda\mu\beta}=-J_{\mu\lambda\beta}\equiv
\nabla_{\lambda}R_{\mu\beta}-\nabla_{\mu}R_{\lambda\beta}$.
Notice that because of (\ref{belsym}) this is the only independent
divergence of the Bel tensor.
The fundamental result we have from (\ref{divbel}) is that $B$ is
divergence-free when the `current' source of matter $J_{\lambda\mu\beta}$
vanishes. This includes
all Einstein spaces (where $R_{\mu\nu}=\Lambda g_{\mu\nu}$), so that
in particular this implies that the Bel-Robinson tensor, which is
just the specialization of the Bel tensor for vacuum, is divergence-free.
The divergence-free property of $B$ in these cases allows us to obtain
conserved currents in the same way as it is usally done using the
energy momentum tensor, once there exist Killing vector fields
in our spacetime, as we will presently see.

These conserved currents are nothing but divergence-free vector fields,
so that conserved quantities can be obtained by means of the Gauss
theorem integrating over appropiate domains of the manifold $\varietat$
\cite{HE}. Following \cite{jose.s-e} one can define the
current related to the Bel tensor with respect to three arbitrary
Killing vector fields $\vec \xi^{}_1$, $\vec \xi^{}_2$, $\vec \xi^{}_3$
as 
\begin{equation}
  \label{eq:current}
  j_{\mu}\left(R_{[2],[2]};
\vec{\xi}^{}_1,\vec{\xi}^{}_2,\vec{\xi}^{}_3\right)\equiv
B_{(\alpha\beta\lambda)\mu}\,\xi^{\alpha}_1\xi^{\beta}_2\xi^{\lambda}_3=
B_{(\alpha\beta\lambda\mu)}\,\xi^{\alpha}_1\xi^{\beta}_2\xi^{\lambda}_3.
\end{equation}
The divergence of this current can be computed to give
\begin{eqnarray}
\nabla_{\mu}j^{\mu}\left(R_{[2],[2]};
\vec{\xi}^{}_1,\vec{\xi}^{}_2,\vec{\xi}^{}_3\right)&=&
\nabla^\mu B_{(\alpha\beta\lambda\mu)}
\,\xi^{\alpha}_1\xi^{\beta}_2\xi^{\lambda}_3 +
3 B_{(\alpha\beta\lambda\mu)}
\nabla^{(\mu}_{ } \xi^{\alpha)}_{(1}\xi^{\beta}_2\xi^{\lambda}_{3)}\nonumber\\
&&=\nabla^\mu B_{(\alpha\beta\lambda\mu)}
\,\xi^{\alpha}_1\xi^{\beta}_2\xi^{\lambda}_3, \nonumber
\end{eqnarray}
using the fact that
$\vec\xi_{A\,(\alpha;\beta)}=0$ ($A=1,2,3$)
Therefore, the vanishing of the divergence of the Bel tensor
implies the vanishing of that of $\vec j\left(R_{[2],[2]};
\vec{\xi}^{}_1,\vec{\xi}^{}_2,\vec{\xi}^{}_3\right)$, which constitutes
then a conserved current.

But the divergence of the Bel tensor does not vanish in general.
As mentioned in the Introduction, these currents, if not
conserved, lead to the description of the interchange of some quantities
between different physical systems,
because it is the {\it total} quantity defined for the whole system
which is indeed conserved \cite{jose.s-e}. The interchange
of these quantities, constructed from the currents for the
so-called super-energy
tensors (s-e quantites, then) for the case of the coupling
between the scalar and gravitational fields has been already shown in
\cite{jose.s-e}, see also \cite{contri},
and explicit examples are given in \cite{rjr}, were neither $\vec j$ nor
the current associated with the s-e tensor of a
scalar field $\phi$, $\vec j_{\phi}$, are conserved, but
the mixed current $\vec j+\vec j_{\phi}$ is indeed divergence-free. 

These examples of interchange of s-e, though, are to be found
in spacetimes admitting a $G_2$ group of isometries acting on
spacelike surfaces but not orthogonally transitively, the reason
being that otherwise the s-e currents are independently conserved.
This is due to the fact that the currents obtained from the
s-e tensor for the gravitational field (Bel tensor) are
conserved automatically depending on some geometrical properties of the
Killing vectors used in its construction, independently of the form
of the Ricci tensor and thus of the matter content. 
This property, far from being undesired, is what one would expect
from a good generalization of the energy-momentum tensor
Indeed, the properties we present later in section \ref{sec:prop}
that lead to the
vanishing of the divergenge of $\vec j \left(R_{[2],[2]};
\vec{\xi}^{}_1,\vec{\xi}^{}_2,\vec{\xi}^{}_3\right)$ are the
analogous to some very well known statements
involving both the Ricci tensor,
and hence the energy-momentum tensor, and two (or one) Killing vector fields
generating a $G_2$ or $G_1$ group of isometries
(see \cite{papapetrou,kundttrump,carter69,chuscomm,carter72,KRAM}).
Let us recall them altogether in the following theorem \ref{teo:orthtran},
just after giving some remarks.

\section{Geometric properties of the energy-momentum currents}
We say that a vetor field $\vec v$ is
`integrable' (or hypersurface orthogonal)
when $\bm v \wedge \d \bm v=0$
(in components $v_{[\alpha}v_{\beta;\gamma]}=0$), and second,
two non-null vector fields orthogonal to two given vectors
$\vec v$ and $\vec w$ generate surfaces whenever the
two 1-forms $\bm v$ and
$\bm w$ associated to the vector fields $\vec v$ and $\vec w$ satisfy
$\bm v \wedge \bm w \wedge \d \bm w=
\bm v \wedge \bm w \wedge \d \bm v=0$.

When $\vec \xi$ and $\vec \eta$ are two non-null
Killing vector fields, the group $G_2$ generated by them
is said to act orthogonally transitively
when the vector fields orthogonal to the group orbits
generate surfaces. This means that
$\bm \xi \wedge \bm \eta \wedge \d \bm \eta=
\bm \xi \wedge \bm \eta \wedge \d \bm \xi=0$, which in components read
$\xi^{}_{[\alpha} \eta^{}_\beta \eta^{}_{\lambda;\rho]}=0,\eta^{}_{[\alpha} \xi^{}_\beta \xi^{}_{\lambda;\rho]}=0.$

Defining the currents asociated with the usual
energy-momentum tensor
$T_{\alpha\beta}$, which equates the Einstein tensor
$S_{\alpha\beta}=R_{\alpha\beta}-\frac{1}{2} R\; \g_{\alpha\beta}$
via the Einstein field equations, with respect to a given Killing vector field
$\vec \xi$, $\vec\CT{\xi}{}$, as
$
\CT{\xi}{\alpha}\equiv \xi^\beta T_{\alpha\beta},
$ 
one can present the known results above mentioned as follows:

\begin{theorem}\cite{papapetrou,kundttrump,carter69,chuscomm,carter72,KRAM}
Let $\vec \xi$ be a non-null Killing
vector field and $\vec\CT{\xi}{}$ its energy-momentum tensor current.
If $\vec \xi$ is integrable, then
$\CT{\xi}{[\beta}\xi^{}_{\lambda]}=0$.

If the spacetime admits two independent non-null Killing
vector fields $\vec \xi$ and $\vec \eta$, let
$\vec\CT{\xi}{}$ and $\vec\CT{\eta}{}$ be
their respective energy-momentum tensor currents.
If $[\vec \xi,\vec \eta]=0$ and the group acts
orthogonally transitively, then
$\CT{\xi}{[\beta}\xi^{}_\lambda\eta^{}_{\mu]}=
\CT{\eta}{[\beta}\xi^{}_\lambda\eta^{}_{\mu]}=0$.
\label{teo:orthtran}
\end{theorem}

\section{Geometrical properties of the Bel currents}
\label{sec:prop}
Following (\ref{eq:current}), when our spacetime admits
two independent Killing vector fields
$\vec\xi$ and $\vec\eta$, there appear 4 different Bel currents
associated with them.
Labeling our two Killing vectors as
$\vec\xi_L$ (where $L,M,N=1,2$) such that $\vec\xi_1\equiv \xi$,
$\vec\xi_2 \equiv \vec\eta$, the associated Bel currents are expressed as
\be
\vec\CB{\xi_L}{\xi_M}{\xi_N}{}{}\hspace{1cm}
(=\vec\CB{\xi_{(L}}{\xi_M}{\xi_{N)}}{}{}).
\label{eq:4currents}
\ee
Of course, if our spacetime admits only a Killing vector field
$\vec\xi$, then its associated Bel current is
given by (\ref{eq:4currents}) with $L=M=N=1$ and $\vec\xi_1\equiv \vec\xi$.

\begin{theorem}
Let $\vec \xi$ be a non-null Killing
vector field.
If $\vec \xi$ is integrable, then
$\CB{\xi}{\xi}{\xi}{[\beta}{}\xi^{}_{\lambda]}=0$,

If the spacetime admits two independent non-null Killing
vector fields $\vec \xi$ and $\vec \eta$, let (\ref{eq:4currents})
be their associated Bel currents.
If $[\vec \xi,\vec \eta]=0$ and the group acts
orthogonally transitively, then
\be
\CB{\xi_L}{\xi_M}{\xi_N}{[\beta}{}\xi^{}_\lambda\eta^{}_{\mu]}=0.
\label{eq:fi1}
\ee
\label{teo:belcurrents}
\end{theorem}
Equations (\ref{eq:fi1}) hold if and only if
\be
\CB{\xi_L}{\xi_M}{\xi_N}{}{\alpha}=a_{LMN}(x^\beta)\xi^\alpha+
b_{LMN}(x^\beta)\eta^\alpha,
\label{eq:prop}
\ee
where the functions $a$'s ($a_{(LMN)}=a_{LMN}$)
and $b$'s ($b_{(LMN)}=b_{LMN}$) must satisfy some relations involving
the norms and the product of the Killing vectors to account for the
symmetric character of the Bel tensor.
But more importantly, taking the Lie derivative
with respect to both Killing vector fields of equation (\ref{eq:prop}),
we have that
$$
\lie_{\vec\xi_P}\left(B^\alpha{}_{(\beta\lambda\mu)}\xi^\beta_L
\xi^\lambda_M\xi^\mu_N \right)=\xi^\rho_P\nabla_\rho\left(a_{LMN}\right)
\xi^\alpha+\xi_P^\rho\nabla_\rho\left(b_{LMN}\right)
\eta^\alpha,
$$
which, since the Bel tensor is invariant under the action of isometries
by construction and the group is Abelian and thus
the left hand side of the equation vanishes, leads then to
$$
\xi^\rho_P\nabla_\rho\left(a_{LMN}\right)=
\xi_P^\rho\nabla_\rho\left(b_{LMN}\right)=0
$$
for $P=1,2$. Taking now the divergence of equation 
(\ref{eq:prop}) we have thus
$$
\nabla_\rho \CB{\xi_L}{\xi_M}{\xi_N}{}{\rho}=
\xi^\rho_P\nabla_\rho\left(a_{LMN}\right)+
\xi_P^\rho\nabla_\rho\left(b_{LMN}\right)=0.
$$

The cases when $\vec \xi$ and/or $\vec\eta$ is integrable can be treated
a special cases of equation (\ref{eq:prop}) with $LMN=111$ with
$b_{111}=0$ and/or $LMN=222$ with $a_{222}=0$ respectively.
All these results can be summarized as follows:
\begin{corollary}
Let $\vec \xi$ be a non-null Killing
vector field.
If $\vec \xi$ is integrable then its corresponding Bel current
$\vec\CB{\xi}{\xi}{\xi}{}{}$ is divergence-free.

If the spacetime admits two independent non-null Killing
vector fields $\vec \xi$ and $\vec \eta$, let
(\ref{eq:4currents}) be their associated Bel currents.
If $[\vec \xi,\vec \eta]=0$ and the group acts
orthogonally transitively, then the four coresponding Bel currents
(\ref{eq:4currents}) are divergence-free.

\end{corollary}

\section*{Acknowledgments}
The authors thank financial support from the Basque Country 
University under project number UPV 172.310-G02/99.
RL is much indebted to J.A. Valiente-Kroon for enlightening
discussions, and also acknowledges the finantial support
of the LOC and of the Basque Government under fellowship BFI98.79.
JMMS and RV thank financial support from the Generalitat de
Catalunya, and from the Ministerio de Educaci\'on y Cultura,
refs. 98SGR 00015 and PB96-0384, respectively.
RV also thanks the Spanish Secretar\'{\i}a de Estado
de Universidades, Investigaci\'on y Desarrollo, grant No. EX99 52155527.

\end{document}